\documentclass[%
 preprint,
 amsmath,amssymb,
 aps,
pra,
]{revtex4-2}
\usepackage{soul,xcolor}
\usepackage{multirow} 
\usepackage{graphicx}
\usepackage{dcolumn}
\usepackage{bm}
\usepackage{float}

\usepackage[colorlinks,
            urlcolor=blue,
            linkcolor=blue,
           anchorcolor=blue,
            citecolor=blue]{hyperref}

\begin{document}

\setstcolor{red}


\title{Acoustic helical dichroism enhanced by chiral quasi-bound states in the continuum}

\author{Qing Tong}
\affiliation{Department of Physics, City University of Hong Kong, Tat Chee Avenue, Kowloon, Hong Kong, China}%
\author{Tong Fu}
\affiliation{Department of Physics, City University of Hong Kong, Tat Chee Avenue, Kowloon, Hong Kong, China}%
\author{Yuqiong Cheng}
\affiliation{Department of Physics, City University of Hong Kong, Tat Chee Avenue, Kowloon, Hong Kong, China}%
\author{Shubo Wang}\email{shubwang@cityu.edu.hk}
\affiliation{Department of Physics, City University of Hong Kong, Tat Chee Avenue, Kowloon, Hong Kong, China}


\date{\today}
            
\begin{abstract}
Acoustic helical dichroism (HD) arises from the interaction between vortex beams carrying orbital angular momentum (OAM) and chiral media, yet such chiral sound-matter interactions are typically weak. Here, we employ quasi-bound states in the continuum (QBICs) in acoustic meta-cavities composed of coupled Helmholtz resonators to enhance acoustic HD. We design both achiral and chiral meta-cavities that support QBICs in the form of vortex states with high Q-factors. Using full-wave numerical simulations, we show that the QBICs in the achiral meta-cavities cannot enhance acoustic HD due to the absence of a chiral wavefront. In contrast, the chiral meta-cavity exhibits a pronounced HD enhancement through the QBICs with a 3D helical wavefront, which can be excited by incident waves either with or without OAM. Our work identifies two essential requirements for enhancing acoustic HD effect via QBICs: a high Q-factor of the states and 3D chirality of the state fields, which usually compromise each other in conventional acoustic resonators. The findings open new avenues for achieving strong chiral sound-matter interactions, with potential applications in acoustic chiral sensing and acoustic OAM manipulation.
\end{abstract}

\maketitle
\newpage

\section{\label{sec:level1}INTRODUCTION}

Chirality is a fundamental concept that describes the property of an object that cannot be superimposed on its mirror image with any translation or rotation. It plays a pivotal role in wave-matter interactions and underpins various intriguing chiral-optical and chiral-acoustic phenomena \cite{r1, r2, r3, r4, r5, r6, r7}. Vortex beams carrying intrinsic orbital angular momentum (OAM) exhibit chiral characteristics through their helical wavefronts. The vortex beams with opposite OAM can interact with chiral matter and give rise to different responses in transmission, reflection, and absorption, a phenomenon known as the helical dichroism (HD). This phenomenon arises from the chiral wave-matter interaction that directly depends on the handedness of the vortex beams. The HD has been well studied in optics both theoretically \cite{r8, r9, r10, r11} and experimentally \cite{r12, r13, r14, r15, r16}. Despite the extensive research on acoustic vortices \cite{r17, r18, r19, r20, r21, r22}, the mechanisms and properties of acoustic HD have not been thoroughly investigated.

Akin to its optical counterpart, acoustic HD suffers from weak chiral wave-matter interactions. To address this problem, several approaches have been proposed, such as employing symmetry breaking and non-Hermitian exceptional points \cite{r23, tong2023, r24, r25}. An alternative and highly promising approach is to employ quasi-bound states in the continuum (QBICs), which exhibit high Q-factors and long lifetimes that can significantly enhance wave-matter interactions. This approach has been well established in optics \cite{r26, r27, r28}, enabling dramatic enhancement of both intrinsic \cite{r29, r30} and extrinsic \cite{r31, r32} chiral light-matter interactions. In contrast, acoustic research on bound states in the continuum (BICs) and QBICs has primarily focused on their realization and observation in various structures, such as in open acoustic resonators \cite{r33, r34, r35, r36, r37} and phononic crystal slabs \cite{r38}, leaving their applications in chiral sound-matter interactions largely unexplored. 

In this study, we apply full-wave simulations to investigate the influence of acoustic QBICs on the HD effect. We design acoustic \textcolor{black}{QBIC} meta-cavities using multiple Helmholtz resonators coupled through slit channels. The meta-cavities have $C_4$ rotational symmetry and support vortex QBICs with high Q-factors, which exhibit strong field confinement at the center of the meta-cavities. Both achiral and chiral meta-cavities are considered for comparison. The achiral meta-cavity supports vortex QBICs with a planar wavefront and induces negligible HD. In contrast, the chiral meta-cavity supports vortex QBICs with a helical wavefront, which can be excited by incident waves either with or without OAM. These vortex QBICs with helical wavefronts can give rise to dramatically enhanced HD effects. The results demonstrate that acoustic vortex QBICs, despite having high Q-factors, may not induce strong chiral sound-matter interactions due to their standing-wave nature. A necessary condition is that these states must have helical wave fields with intrinsic 3D chirality. 

\section{\label{sec:level2}QBICS IN ACOUSTIC META-CAVITIES}

\begin{figure}[t]
\centering
\includegraphics[scale=1]{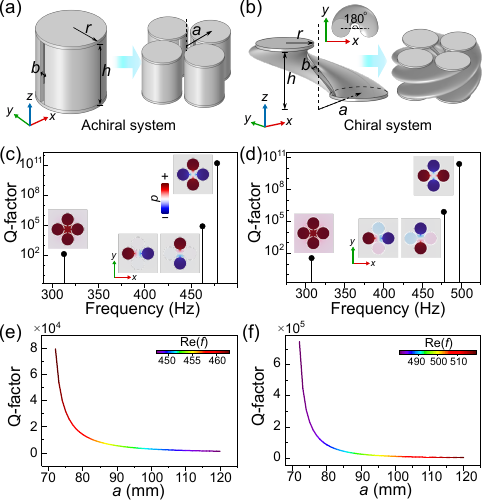}
\caption{(a) Achiral meta-cavity composed of four identical Helmholtz resonators with slit openings. (b) Chiral meta-cavity composed of four identical twisted Helmholtz resonators with slit openings. Each resonator of the achiral and chiral meta-cavities has radius $r=50$ mm, height $h=100$ mm, slit width $b=10$ mm, and radial distance $a$ from the central axis. All resonators have a shell thickness $t=5$ mm. The inset in (b) indicates the twist angle of 180° for each resonator. Eigenfrequencies and Q-factors for the lowest four eigenmodes of (c) the achiral meta-cavity and (d) the chiral meta-cavity with $a=72$ mm. The insets show the eigen pressure field on a cross section of the meta-cavities. The Q-factor of dipole modes as a function of the radial distance $a$ for (e) the achiral meta-cavity and (f) the chiral meta-cavity. Line colors indicate the real part of the eigenfrequency. } \label{fig1}
\end{figure}

We consider the achiral acoustic meta-cavity shown in Fig. \ref{fig1}(a), which comprises four cylindrical Helmholtz resonators arranged on a circle of radius $a$. Each resonator has radius $r$, height $h$, and a slit of width $b$. The slits lead to radiation loss of the resonators, and the four resonators can couple with each other through the slits. In addition to the achiral meta-cavity, we consider the chiral meta-cavity shown in Fig.\ref{fig1}(b), which consists of four twisted Helmholtz resonators arranged on a ring of radius $a$. Here, each resonator is obtained by twisting the Helmholtz resonator in Fig.\ref{fig1}(a) by 180 degrees with respect to the center axis of the meta-cavity. The geometric parameters $a, b, h$, and $r$ take the same values as in the case of Fig. \ref{fig1}(a). Both the chiral and achiral meta-cavities have a $C_4$ rotational symmetry. Hard boundary conditions are set on all the boundaries of the chiral and achiral meta-cavities.

We conduct numerical simulations by using a finite-element-method package \cite{r39} to determine the eigenmodes of the meta-cavities. Figure \ref{fig1}(c) shows the eigenmodes and their Q-factors supported by the achiral meta-cavity, where the insets show the pressure field on the $xy$ symmetry plane. We notice that four eigenmodes exist in the considered frequency range [300 Hz, 500 Hz]: a monopole mode, two dipole modes, and a quadrupole mode. The dipole and quadrupole modes are QBICs arising from the destructive interference of the Helmholtz resonators. Their Q-factors remain finite due to the radiation loss, i.e., the field leakage via the slits in the cavities. The two dipole modes are degenerate and orthogonal. By exciting the two dipole modes with a phase difference of $\pi/2$, an acoustic vortex state carrying intrinsic OAM can be obtained in the meta-cavity, corresponding to a vortex QBIC. Notably, this vortex state exhibits no phase variation along $z$ direction, i.e., the vortex fields carry pure 2D chirality, as we will show later. 

Figure \ref{fig1}(d) shows the eigenmodes and their Q-factors supported by the chiral meta-cavity in Fig. \ref{fig1}(b). Similar to the achiral meta-cavity, the chiral meta-cavity supports a monopole mode, two dipole modes, and a quadrupole mode in the considered frequency range. The two dipole modes remain degenerate, as protected by the $C_4$ rotational symmetry. Notably, each dipole mode exhibits a helical field pattern with an effective axial wavenumber $k_z=\pi/h$  due to the twisting feature of the meta-cavity. Consequently, when the two dipole modes are excited with a phase difference of $\pi/2$, their superposition gives rise to a vortex QBIC carrying an additional \textcolor{black}{effective} phase factor $e^{ik_z z}$ \textcolor{black}{within the chiral meta-cavity}, compared to the vortex QBIC in the achiral meta-cavity. The helical field pattern of the vortex QBIC features non-zero 3D chirality that is essential to achieve acoustic HD. 

 The Q-factors and eigenfrequencies of the QBICs strongly depend on the coupling among the four Helmholtz resonators. To understand this dependence, we simulate the Q-factors and eigenfrequencies of the dipole QBICs as a function of the parameter $a$ (i.e., radial distance between the Helmholtz resonators and the center axis of meta-cavities). The results are presented in Figs. \ref{fig1}(e) and \ref{fig1}(f) for the achiral and chiral meta-cavities, respectively, where the color of the lines denotes the real part of eigenfrequency Re$(f)$. It is seen that both Re$(f)$ and Q-factor decrease with the increase of $a$. This can be understood as follows. The monopole and dipole modes in Fig.\ref{fig1}(c) correspond to the “bonding” and “anti-bonding” of opposing Helmholtz resonators, respectively. Their eigenfrequency difference is proportional to the coupling strength between the Helmholtz resonators. Increasing the radial distance $a$ will weaken the coupling and thus reduce the dipole mode frequency. Meanwhile, increasing $a$ will reduce the destructive interference of the Helmholtz resonators and increase the radiation loss, thus reducing the Q-factor of the dipole QBICs.  For the dipole QBICs in the chiral meta-cavity, Fig.\ref{fig1}(f) shows that the eigenfrequency Re$(f)$ and Q-factor exhibit different dependence on the radial distance $a$. A larger $a$ leads to a smaller Q-factor but a larger Re$(f)$. The smaller Q-factor can be attributed to the larger radiation loss, similar to the achiral case. The larger Re$(f)$ is attributed to the simultaneous variations in the volume, slit opening, and coupling of the Helmholtz resonators. Increasing $a$ increases the volume and reduces the coupling, both of which lower the eigenfrequency. However, it also increases the area of the slit openings, which increases the eigenfrequency \cite{r40, r41}.

\section{\label{sec:level3}ACOUSTIC HD IN ACHIRAL META-CAVITY}

\begin{figure}[htb]
\centering
\includegraphics[scale=1]{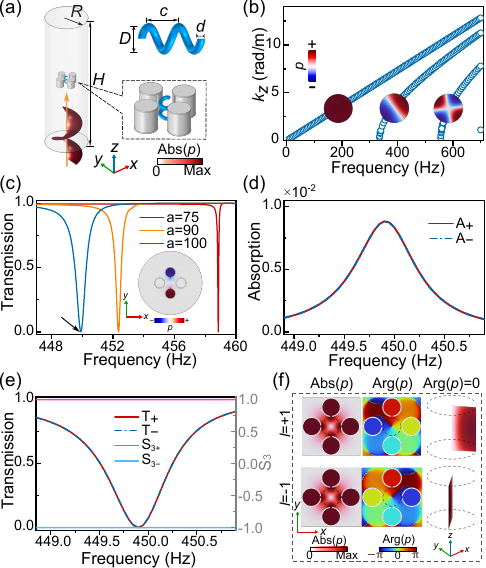}
\caption{(a) Achiral meta-cavity system under the incidence of a vortex guided wave. The meta-cavity is located inside a cylindrical waveguide of radius $R=300$ mm and height $H=2000$ mm. A helix particle (with $D=40$ mm, $d=16$ mm, and $c=50$ mm) is placed at the center of the meta-cavity, as shown by the zoom-in. (b) Three lowest-order guided modes of the cylindrical waveguide. (c) Waveguide transmission for the achiral meta-cavities with $a = 75$ mm, $90$ mm, and $100$ mm, under the incidence of the linear dipole guided wave $p_y$. The inset shows the excited pressure field inside the meta-cavity. (d) Absorption of the helix particle under the incidence of vortex guided waves with $l = \pm1$, where we set $a = 100$ mm. (e) Waveguide transmission (solid red and dashed blue lines) for the achiral meta-cavity ($a = 100$ mm) without the helix, under the incidence of vortex guided waves with $l = \pm1$. The solid magenta and blue lines denote the normalized Stokes parameters $S_3$ of the local velocity field at the center of the meta-cavity. (f) Pressure amplitude and phase of the excited vortex QBICs inside the meta-cavity. The dashed black circles in the profiles of Abs($p$) and Arg($p$) indicate the location of the wavefront Arg($p$) = 0.} \label{fig2}
\end{figure}

To investigate how the vortex QBICs in the achiral meta-cavity influence acoustic HD, we employ the system shown in Fig. \ref{fig2}(a), where the meta-cavity is put inside a cylindrical waveguide with radius $R = 300$ mm and length $H = 2000$ mm. The vortex QBICs with opposite topological charge $l=\pm1$ (corresponding to opposite OAM), excited by the vortex guided waves, can interact with a lossy helix particle located at the cavity center and give rise to different absorption. The helix has major diameter $D$, minor diameter $d$, and pitch $c$.  The loss of the helix is characterized by the imaginary part of sound speed $v_0=343(1+i\alpha)$ m/s. Figure \ref{fig2}(b) shows the three lowest-order guided waves and their dispersion relations in the cylindrical waveguide. The cut-off frequency of the dipole guided wave is about 335 Hz, which is below the eigenfrequency of the dipole modes in the meta-cavity. Thus, the dipole guided wave can excite the dipole QBICs in the meta-cavity. To verify this, we simulate the transmission of the dipole guided wave in the presence of the meta-cavity (without the helix particle). The results are shown in Fig. \ref{fig2}(c) for the meta-cavities with $a = 75$ mm, $90$ mm, and $100$ mm. As noticed, a dip appears in the transmission spectra, corresponding to a resonance at which the transmission vanishes, and the dipole guided wave is completely reflected. The Q-factor of this resonance, as indicated by the half-height width of these dips, reduces as $a$ increases, consistent with the results in Fig. \ref{fig1}(e). The inset in Fig. \ref{fig2}(c) shows the pressure field inside the meta-cavity at the resonance frequency for the case $a = 100$ mm, confirming the excitation of the dipole QBIC. These results demonstrate that the dipole guided wave indeed can excite the dipole QBIC in the achiral meta-cavity. Therefore, an incident vortex guided wave, corresponding to the superposition of two orthogonal dipole guided waves with a phase difference of $\pi/2$, can excite the vortex QBICs in the achiral meta-cavity, inducing strong interaction with the chiral particle. Figure \ref{fig2}(d) shows the absorption of the incident vortices with opposite topological charge $l=\pm1$, where $A_+ (A_-)$ denotes the absorption of the vortex with $l=+1 (l=-1)$, where we set $\alpha=0.05$ for the helix absorption. We notice that they exhibit the same absorption in the considered frequency range. This demonstrates that the vortex QBICs in the achiral meta-cavity cannot induce acoustic HD, despite their chiral characteristic associated with the OAM.

To understand the mechanism behind the vanished HD in the achiral meta-cavity, we remove the chiral particle and simulate the transmission of the vortex guided waves with $l =\pm1$. The results are denoted by the solid red and dashed blue lines in Fig. \ref{fig2}(e). As seen, the transmission vanishes at the frequency of the vortex QBICs. Figure \ref{fig2}(f) shows the corresponding pressure amplitude Abs($p$) and phase Arg($p$) on the $xy$ plane, confirming the excitation of the vortex QBICs with $l =\pm1$. To characterize the \textcolor{black}{chiral property} of the vortex QBICs, we evaluate the normalized Stokes parameter $S_{3} =\frac{2 \operatorname{Im}\left(v_{x} v_{y}^{*}\right)}{\left|v_{x}\right|^{2}+\left|v_{y}\right|^{2}}$ based on the velocity field $\mathbf{v}=\left(v_x, v_y\right)$ at the center of the meta-cavity, which is directly related to the topological charge $l$ of the vortices\cite{wang2025a, liu2025}. We notice that the normalized Stokes parameter for the vortex QBIC with $l = 1$ maintains a constant value of $S_{3+}=+1$ (denoted by the solid magenta line) in the considered frequency region [449 Hz, 451 Hz]. In contrast, the normalized Stokes parameter for the vortex QBIC with $l = -1$ maintains the constant value of $S_{3-} =-1$ (denoted by the solid blue line). Thus, the vortex QBICs with $l =\pm1$  \textcolor{black}{exhibit opposite spatial chirality near the resonance frequency.} Importantly, this chirality is intrinsically 2D because the fields of the vortex QBICs are invariant along $z$ direction, i.e., the chirality is entirely confined to the 2D transverse plane, as confirmed by the pressure wavefronts Arg($p$) = 0 in Fig. \ref{fig2}(f). Therefore, the vanished HD in the achiral meta-cavity can be attributed to the 2D chiral nature of the vortex QBICs.

\section{\label{sec:level4} ACOUSTIC HD IN CHIRAL META-CAVITY}    

\begin{figure}[h]
\centering
\includegraphics[scale=1]{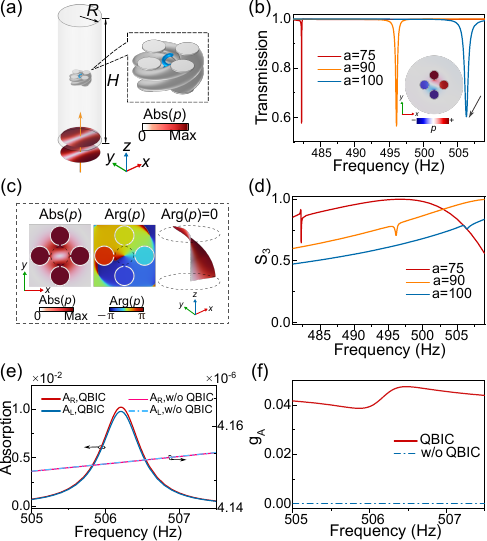}
\caption{(a) Chiral meta-cavity system under the incidence of a linear dipole guided wave. The waveguide and helix particle are the same as in Fig. \ref{fig2}(a). (b) Waveguide transmission for the chiral meta-cavities with radii $a = 75$ mm, $90$ mm, and $100$ mm, under the incidence of the linear dipole guided wave polarized along $y$-direction. (c) Pressure amplitude and phase of the excited vortex QBIC. The dashed circles marked in the profiles of Abs($p$) and Arg($p$) indicate the location of the wavefront Arg($p$) = 0. (d) Normalized Stokes parameter $S_3$ of the velocity field at the center of the meta-cavities with radii $a = 75$ mm, $90$ mm, and $100$ mm. (e) Absorption of the left-handed and right-handed helices for the systems with and without the chiral meta-cavity. We set $a = 100$ mm for this case. (f) Absorption dissymmetry factor $g_A$ for the systems with and without the chiral meta-cavity, corresponding to the results in (e).} \label{fig3}
\end{figure}

To investigate how the vortex QBICs in the chiral meta-cavity influence acoustic HD, we employ the system shown in Fig. \ref{fig3}(a), where the chiral meta-cavity is put inside the same cylindrical waveguide as in Fig.  \ref{fig2}(a). In this case, we excite the vortex QBICs using the linear dipole guided wave (velocity polarized along $y$ direction) and determine the differential absorption induced by two lossy helix particles of opposite handedness. The helices have the same geometric parameters as shown in Fig.\ref{fig2}(a). To confirm the excitation of the vortex QBICs, we simulate the transmission of the incident dipole guided wave for the chiral meta-cavities with different radii $a = 75$ mm, $90$ mm, and $100$ mm, without the presence of the helix particle. The results are summarized in Fig. \ref{fig3}(b). As noticed, the transmission spectra show resonance dips, and the Q-factor of the resonance reduces as $a$ increases, consistent with the results in Fig. \ref{fig1}(f). Figure \ref{fig3}(c) shows the pressure field on $xy$ plane corresponding to the dip in the case $a = 100$ mm. We notice that the field is localized inside the meta-cavity and exhibits a phase singularity in the center, confirming the excitation of the vortex QBIC. Importantly, the field exhibits 3D chirality, as shown by the wavefront Arg($p$)=0. To understand the \textcolor{black}{chiral property} of the excited vortex QBIC, we calculate the normalized Stokes parameter $S_3$ based on the velocity field in the center of the meta-cavity. The results are shown in Fig. \ref{fig3}(d). As seen, the normalized Stoke parameter remains in the range $0.5< S_3 <1$ for the considered three cases, corresponding to an elliptically polarized velocity field. This demonstrates that the incident dipole guided wave predominantly excites the vortex QBIC with $l=+1$, as shown in Fig. \ref{fig3}(c). We note that the incident guided wave also excites the opposite vortex QBIC with a much smaller amplitude, which is attributed to the breaking of mirror symmetry by the chiral meta-cavity. 

We further investigate the HD effect by simulating the absorption spectra of the system in Fig. \ref{fig3}(a), with embedded lossy helices of opposite handedness. The results are shown in Fig.  \ref{fig3}(e) (solid red and blue lines) for the meta-cavity with radius $a = 100$ mm, where $A_R$ and $A_L$ denote the absorption to the right-handed and left-handed helices, respectively. We notice that the absorption is significantly enhanced at the frequency of the QBIC, reaching the value of $1\times10^{-2}$. Importantly, the absorption $A_R$ and $A_L$ are different at the resonance frequency, demonstrating the emergence of acoustic HD effect. For comparison, we remove the chiral meta-cavity and simulate the absorption of the lossy helices excited by an incident vortex guided wave with $l=+1$. The results are denoted by the solid magenta and dashed blue lines in Fig.  \ref{fig3}(e), which shows nearly identical absorption on the order of $10^{-6}$, much smaller than the absorption enhanced by the QBIC. To characterize the HD effect, we further calculate the absorption dissymmetry factor $g_A=2\left|A_R-A_L\right| /\left|A_R+A_L\right|$ for the systems with and without the chiral meta-cavity. As shown in Fig. \ref{fig3}(f), the system with the chiral meta-cavity (with $a=100$ mm) gives rise to a dissymmetry factor $g_A \sim 0.04$. In contrast, the dissymmetry factor is almost zero for the system without the chiral meta-cavity (denoted by the dashed blue line). These results demonstrate the crucial role of the QBIC supported by the chiral meta-cavity, which significantly enhances the chiral sound-matter interaction and thus the acoustic HD effect.

\begin{figure}[!t]
\centering
\includegraphics[scale=1]{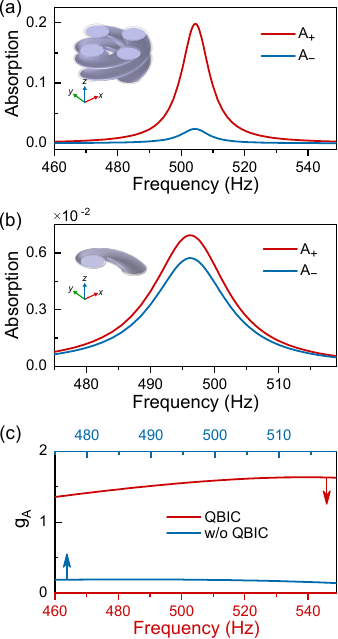}
\caption{(a) Absorption of the lossy chiral meta-cavity under the incidence of vortex guided waves with $l=\pm 1$. We set the sound speed $v_0=343(1+0.01i)$ m/s for the regions (blue colored) inside the four Helmholtz resonators. (b) Absorption of a single lossy Helmholtz resonator under the incidence of vortex guided waves with $l=\pm 1$. (c) Absorption dissymmetry factor $g_A$  for the systems in (a) and (b).} \label{fig4}
\end{figure}

Besides, the chiral \textcolor{black}{QBIC} meta-cavity itself is a chiral structure that can induce differential absorption of the vortex guided waves with $l=\pm1$. To demonstrate this, we consider the system in Fig. \ref{fig3}(a) without the presence of the helix particle. We introduce uniform loss into the four Helmholtz resonators by setting the sound speed to be $v_0=343(1+0.01i)$ m/s. The system is excited by the vortex guided waves with opposite topological charge $l=+1$ and $l=-1$. Figure \ref{fig4}(a) shows the corresponding absorption spectra of the system. We notice that the lossy chiral meta-cavity can induce a large differential absorption of the opposite vortex guided waves, which is strongly enhanced at the frequency of the QBIC and is much larger than that of the lossy helices in Fig. \ref{fig3}(e). \textcolor{black}{This is because that the chiral meta-cavity can strongly confine fields inside the lossy region due to the QBIC, and the structure is geometrically larger than the helices.}  For comparison, we also simulate the absorption spectra induced by one single Helmholtz resonator, as shown by the inset in Fig. \ref{fig4}(b), which does not support a QBIC in the considered frequency range. In this case, the differential absorption is about two orders of magnitude smaller than in the case of the chiral meta-cavity. Figure \ref{fig4}(c) shows the dissymmetry factor $g_A$ for the two cases. As seen, the chiral meta-cavity with QBIC (solid red line) induces a much larger dissymmetry factor than the single Helmholtz resonator (solid blue line). These results further demonstrate the strong enhancement of acoustic HD by the QBIC in the chiral meta-cavity.

\section{\label{sec:level5} CONCLUSION }

In conclusion, we investigate the acoustic HD effect in acoustic meta-cavities supporting chiral QBICs (i.e., vortex QBICs). Two types of meta-cavities are considered: the achiral meta-cavity composed of cylindrical Helmholtz resonators and the chiral meta-cavity composed of twisted Helmholtz resonators. Both meta-cavities support vortex QBICs that can be excited by guided waves. By putting small helix particles inside the meta-cavities and simulating the absorption spectra, we find that the vortex QBICs in the achiral meta-cavity cannot induce the HD effect despite the chirality associated with OAM. This can be attributed to the absence of 3D chirality in the vortex fields of the achiral meta-cavity. In contrast, the vortex QBIC in the chiral meta-cavity exhibits a strongly enhanced HD, compared to the HD induced by vortex guided waves without the meta-cavity. We further demonstrate that the chiral meta-cavity alone can induce strong acoustic HD, which is much stronger than the HD induced by a single chiral Helmholtz resonator. 

Our work uncovers two essential requirements for enhancing chiral sound-matter interactions through QBICs: a high Q-factor of the states and 3D chirality of the state fields, which usually compromise each other in conventional acoustic resonators.  While the non-chiral acoustic resonators can give rise to 2D chiral fields (e.g., plane vortices), they generally cannot give rise to 3D chiral fields due to their localized standing wave nature. For chiral acoustic resonators, the structural chirality induces 3D “twisting” of the fields, giving rise to an effective wavevector along the twisting axis despite the standing wave nature of the fields. The roles of 2D and 3D chirality of acoustic fields in this work are similar to the roles of optical spin and helicity in chiral light-matter interactions. The results provide a new mechanism for achieving strong chiral sound-matter interactions and can find broad applications in acoustic chiral sensing, OAM manipulation, and acoustic metamaterial design.

\begin{acknowledgments}
The work described in this paper was supported by grants from the National Natural Science Foundation of China (No. 12322416) and Research Grants Council of the Hong Kong Special Administrative Region, China (Project No. AoE/P-502/20).   
\end{acknowledgments}


\bibliography{MyCollection}

\end{document}